\documentclass[twocolumn,preprintnumbers,amsmath,aps]{revtex4-1}

\usepackage{graphicx}
\usepackage{dcolumn}
\usepackage{bm}
\usepackage{color}
\begin{document}
\def\eq#1{(\ref{#1})}
\def\fig#1{Fig.\hspace{1mm}\ref{#1}}
\def\tab#1{\hspace{1mm}\ref{#1}}
\title{Strong-coupling character of superconducting phase in compressed selenium hydride}  
\author{Ewa A. Drzazga-Szcz{\c{e}}{\'s}niak}
\email{ewa.drzazga@pcz.pl}
\author{Adam Z. Kaczmarek}
\affiliation{Institute of Physics, Cz{\c{e}}stochowa University of Technology, Ave. Armii Krajowej 19, 42-200 Cz{\c{e}}stochowa, Poland}
\date{\today}
\begin{abstract}

At present, metal hydrides are considered highly promising materials for phonon-mediated superconductors, that exhibit high values of the critical temperature. In the present study, the superconducting properties of the compressed selenium hydride in its simplest form (HSe) are analyzed, toward quantitative characterization of this phase. By using the state-of-art Migdal-Eliashberg formalism, it is shown that the critical temperature in this material is relatively high ($T_{c}$=42.65 K) and surpass the level of magnesium diboride superconductor, assuming that the Coulomb pseudopotential takes value of $0.1$. Moreover, the employed theoretical model allows us to characterize other pivotal thermodynamic properties such as the superconducting band gap, the free energy, the specific heat and the critical magnetic field. In what follows, it is shown that the characteristic thermodynamic ratios for the aforementioned parameters differ from the predictions of the Bardeen-Cooper-Schrieffer theory. As a result, we argue that strong-coupling and retardation effects play important role in the discussed superconducting state, which cannot be described within the weak-coupling regime.

\end{abstract}
\maketitle
{\bf Keywords:} superconductivity; high-pressure effects; thermodynamic properties; hydrogen-based materials

\section{Introductions}

Among many applications, hydrogen is expected to have potential in building high temperature phononic superconductors \cite{ashcroft1}. In general, one can distinguish two representative phases of superconducting hydrogen with respect to the applied pressure, namely the molecular phase (in the range from $\sim 300$ to $\sim 500$ GPa) \cite{zhang1, cudazzo1, cudazzo2} and the atomic phase (above $\sim 500$ GPa) \cite{yan, mcmahon, zhang2}. In the molecular phase, hydrogen is predicted to yield critical temperature ($T_{C}$) value well above 100 K in the regime of strong electron-phonon coupling $\it i.e.$ the electron-phonon coupling constant ($\lambda$) is expected to be higher than 0.5 \cite{zhang1, cudazzo1, cudazzo2}. On the other hand, above 500 GPa the molecular metallic hydrogen converts into the atomic phase, where the electron-phonon coupling constant can be anomalously high ($\lambda>2$) \cite{yan, mcmahon, zhang2}. 

However, the main downside of using only pure hydrogen in building the high temperature superconductors is the requirement of applying relatively high pressure. It turns out that the implementation of other elements into the hydrogen can considerably reduce such pressure requirements, so that they can be reached under laboratory conditions. This statement is based on earlier predictions of Ashcroft \cite{ashcroft1,ashcroft2}, suggesting that the high temperature phonon-mediated superconducting state may be induced in the hydrogen-rich compounds under high pressure. Specifically, Ashcroft predicts that introduction of the heavier atoms into the metallic hydrogen should allow to lower metallization pressure and still retain relatively high $T_{c}$ values. However, besides many theoretical investigations in this field \cite{szczesniak1}, the scientists have experimentally obtained only several compounds that deserve attention \cite{kim, eremets, drozdov1, drozdov2, drozdov3}. Therefore, further corresponding intensive theoretical and experimental studies are still desirable.

In details, the most interesting representatives of the hydrogen-based compounds are: PtH \cite{kim, szczesniak2, szczesniak3}, SiH$_{4}$ \cite{eremets, szczesniak4}, H$_{3}$S \cite{drozdov1, durajski1}, PH$_{3}$ \cite{drozdov2, durajski2}, and LaH$_{10}$ \cite{drozdov3, kruglov}. Nonetheless, the discovery of superconducting state in sulfur hydride (H$_{3}$S) was the initial breakthrough in the research on the phonon-mediated hydrogen-based superconductors. Specifically, Drozdov {\it et al.} reported record-high, at that time, superconducting critical temperature ($T_{c}$) of 203 K in the ultradense phase of  H$_{3}$S \cite{drozdov1}. As a result, investigations made by Drozdov {\it et al.} motivated further studies on other similar hydrogen-based compounds, aimed at the higher critical temperature values or lower metallization pressures. A natural candidates for such an investigations are hydrides containing other chalcogens, which are isoelectronic to sulfur. In this context, recent studies on elemental chalcogen superconductors suggest that selenium exhibits closest properties of the superconducting state to sulfur, among all chalcogens \cite{zhou, szczesniak5}. Moreover, both corresponding hydrides are strong-coupling and phonon-mediated superconductors with properties strongly dependent on pressure \cite{zhang3}. This facts additionally indicates that Bardeen-Cooper-Schrieffer (BCS) theory \cite{bardeen1, bardeen2} may not be proper in the description of such superconducting phases \cite{carbotte}.

In reference to the above, we provide the theoretical analysis of selected superconducting properties of the selenium hydride superconductor (HSe), the simplest representative of the selenium-based hydride superconductors. Herein, we consider that HSe is exposed to the external pressure of 300 GPa, a level of compression that assures thermodynamic stability of the system and relatively high electron-phonon coupling ($\lambda$=0.79) \cite{zhang3}. Hence, this theoretical study attempts to provided new contribution to the field of research on superconductivity in hydrides. In particular, such approach may result in a much more detailed understanding of the underlying physics of the superconducting state induced in chalcogen hydrides. Due to $\lambda>0.5$, the presented investigations are conducted within the Migdal-Eliashberg formalism \cite{migdal, eliashberg}, that allows to obtain quantitative predictions of the thermodynamics in phonon-mediated superconductors exhibiting high values of the electron-phonon coupling constant \cite{cyrot, carbotte}. For example, mentioned methodology has been already employed in terms of the research on H$_{3}$S systems \cite{durajski1} but also pure chalcogen superconductors (the Se and Te systems \cite{szczesniak5}) with a great results. Specifically, this work concentrates on the temperature-dependent behavior of the pivotal parameters that describes HSe superconductors, such as the superconducting band gap, specific heat or the critical magnetic field.

\section{Theoretical model}

To describe the thermodynamic properties of the selenium hydride at 300 GPa, we employ the Migdal-Eliashberg equations in the isotropic approximation form \cite{migdal, eliashberg, carbotte}. When solving these equation on the imaginary axis, it is possible to determine two pivotal parameters, namely the order parameter ($\Delta_{n}=\Delta\left(i\omega_{n}\right)$) and the wave function renormalization factor ($Z_{n}=Z\left(i\omega_{n}\right)$), given as:
\begin{eqnarray}
\label{eq1}
\Delta_{n}Z_{n} &=& \frac{\pi}{\beta}\\ \nonumber
&\times& \sum_{m=-M}^{M} \frac{K\left(i\omega_{n}-i\omega_{m}\right)-\mu^{\star}\theta\left(\omega_{c}-|\omega_{m}|\right)}
{\sqrt{\omega_m^2Z^{2}_{m}+\Delta_{m}^{2}}}\Delta_{m}^2,
\end{eqnarray}
and
\begin{equation}
\label{eq2}
Z_{n}=1+\frac{1}{\omega_{n}}\frac{\pi}{\beta}\sum_{m=-M}^{M}
\frac{K\left(i\omega_{n}-i\omega_{m}\right)}{\sqrt{\omega_m^2Z^{2}_{m}+\Delta^{2}_{m}}}\omega_{m}Z_{m},
\end{equation}
where $\beta=1\slash k_{B}T$ is the inverse temperature, calculated with respect to the Boltzmann constant ($k_{B}$), that allows us to set the the Matsubara frequency as $\omega_{n}=\left(\pi\slash\beta\right)\left(2n-1\right)$. Moreover, $K\left(z\right) = 2\int_0^{\omega_{\rm{max}}} d \left( \alpha^{2}F(\omega)\omega \right) \left[ (\omega_{n}-\omega_{m})^2+\omega^2 \right]$ and defines the electron-phonon paring kernel. Therein, the $\alpha^{2}F\left(\omega\right)$ denotes the electron-phonon spectral function, conventionally refereed to as the Eliashberg function. Its form is adopted here from the study of Zhang {\it et al.} \cite{zhang3}, where the function was calculated within the linear-response theory via Quantum-ESPRESSO package. To this end, in Eq. (\ref{eq1}), the electron-electron depairing interactions are modeled by the Coulomb pseudopotential ($\mu^{\star}$) parameter, defined as $\mu^{\star}\equiv\mu^{\star}\theta(\omega_{c}-|\omega_{m}|)$, where $\theta$ is the Heaviside function.

The Eliashberg equations on the imaginary axis are solved here by using the self-consistent iterative procedures developed previously in \cite{szczesniak6}. The stability of the numerical procedures is reached at around the 2201 Matsubara frequencies, assuming $T_{0}=4$ K and the phonon frequency cut-off ($\omega_{c}$) equal to $10\omega_{max}$, where $\omega_{\rm{max}}=203.66$ meV and denotes the maximum value of the phonon frequency defined by the adopted $\alpha^{2}F\left(\omega\right)$ function. Finally, $\mu^{\star}$ is set to 0.1, as suggested by Ashcroft for the hydrogen-based superconductors \cite{ashcroft2}.

To improve predictions on the order parameter, the Eliashberg equations are additionally solved on the real axis in the framework of the Pad{\' e} analytic continuation method \cite{beach}. In what follows, the $\Delta\left(\omega\right)$ parameter on the real axis is given as \cite{beach}:
\begin{equation}
\label{eq3}
\Delta\left(\omega\right)=\frac{p_{\Delta 1}+p_{\Delta 2}\omega+...+p_{\Delta r}\omega^{r-1}}
{q_{\Delta 1}+q_{\Delta 2}\omega+...+q_{\Delta r}\omega^{r-1}+\omega^{r}},
\end{equation}
where $p_{\Delta j}$ and $q_{\Delta j}$ denote numerical coefficients, along with $r=550$. In a result we can write the temperature-dependent order parameter as \cite{eliashberg}, \cite{carbotte}:
\begin{equation}
\label{eq4}
\Delta\left(T\right)={\rm Re}\left[\Delta\left(\omega=\Delta\left(T\right),T\right)\right].
\end{equation}
On the basis of Eq. (\ref{eq3}), the superconducting energy band gap at the Fermi level reads: $\Delta_{g}=2\Delta\left(0\right)$, where $\Delta\left(0\right)\simeq\Delta\left(T_{0}\right)$.

\section{Numerical results}

\begin{figure}[hb!]
\includegraphics[width=\columnwidth]{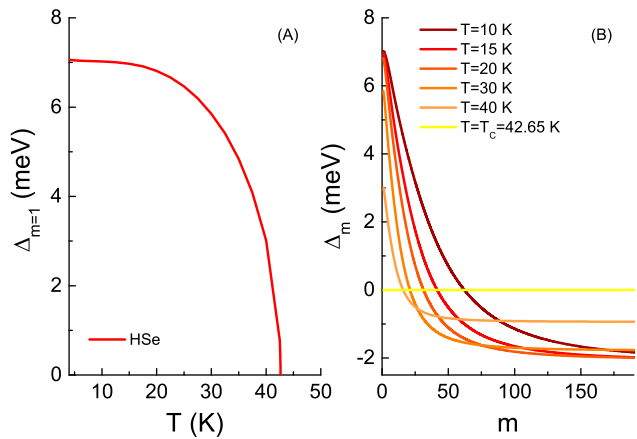}
\caption{(A) The maximum value of the order parameter as a function of the temperature. (B) The order parameter for the selected values of the temperature, as a function of the Matsubara frequencies.} 
\label{f1}
\end{figure}

In the first step, the imaginary-axis solutions of the Eliashberg equations (see Eqs. (\ref{eq1}) and (\ref{eq2})) let us determine the dependence of the maximum value of the order parameter ($\Delta_{m=1}$) on temperature. The corresponding results are depicted in Figure \ref{f1} (A). Here, we remind that the order parameter at 0 K corresponds to the half-width of the energy gap on the Fermi surface. The first thermodynamically stable results are obtained for the temperature $T\geq T_{0}=4$~K. But since this solution still belongs to the characteristic low-temperature plateau, we can assume that $\Delta_{m=1}(4{\rm K})=\Delta_{m=1}(0{\rm K})$. In what follows, we observe that the maximum value of the energy gap in HSe at 300 GPa is equal to $\Delta_{g}=14.12$ meV. In addition to this finding, the Figure \ref{f1} (A) allows us to identify the critical temperature value ($T_{c}$) in the considered superconductor. This is done by arguing the fact that the order parameter takes on the zero value at the critical temperature, marking the superconductor-metal phase transition. In our case, the obtained critical temperature value is $T_{c}=42.65$ K.

To supplement results presented in Figure \ref{f1} (A), in Figure \ref{f1} (B) we plot the order parameter as a function of the Matsubara frequencies, for the selected values of temperature. Note, that all presented functions become practically saturated above 150 Matsubara frequencies and have lorenzian shape. Moreover, with the increase of the temperature, $\Delta_m$ is decreasing, which means that smaller number of the Matsubara frequencies contributes to the solutions of the Eliashberg equations. In this context, results presented in Figure \ref{f1} (B) confirm high accuracy of the calculations conducted initially at 2201 Matsubara frequencies.

\begin{figure}
\includegraphics[width=\columnwidth]{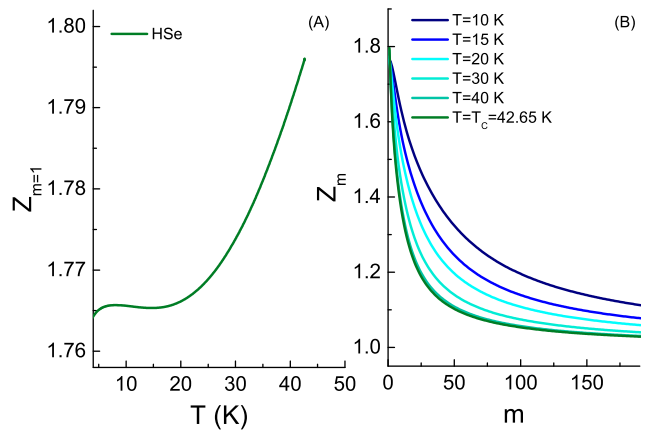}
\caption{(A) The maximum value of the wave function renormalization factor as a function of the temperature; (B) The wave function renormalization factor for the selected values of the temperature, as function of the Matsubara frequencies.} 
\label{f2}
\end{figure}

The above statement on the numerical accuracy is additionally reinforced in the case of the wave function renormalization factor and its temperature-dependent behavior on the imaginary axis, as presented in Figure \ref{f2} (A) and (B). The obtained estimates show that the maximum value of the $Z_{m=1}$ function is slightly increasing when the temperature gets higher. As a result, the wave function renormalization factor behaves similarly to the order parameter in terms of the Matsubara frequencies dependency (see Figures \ref{f1} (B) and \ref{f2} (B)). However, this behavior can be additionally related to the effective mass of electrons ($m_e^*$) through the following relation $m_e^*/m_e\simeq Z_{m=1}$, where electron band mass is denoted by $m_e$. In what follows, the obtained value for the discussed compound is $Z_{m=1}\simeq 1.8$. We note that this value coincides with the estimates calculated from the approximate relation given by $Z_{m=1}=1+\lambda$ (for $\lambda=0.79$ \cite{zhang3}), and again reinforces accuracy of the numerical results obtained here.

\begin{figure*}
\includegraphics[scale=0.5]{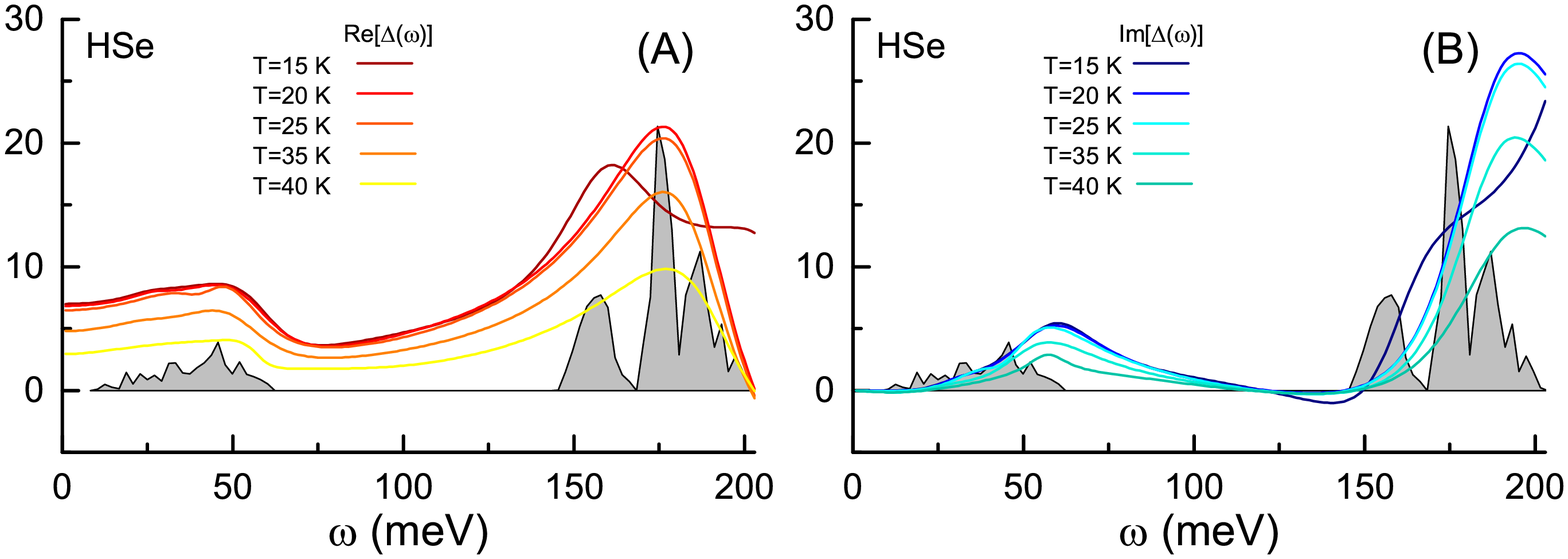}
\caption{(A) The real (A) and imaginary (B) part of the order parameter on the real axis for the selected values of the temperature. The grey shaded area shows the rescaled Eliashberg function, that is adopted in this paper and originally determined in \cite{zhang3}.} 
\label{f3}
\end{figure*}
To determine the precise value of the superconducting energy gap at the Fermi level we conduct the calculations on the real axis. The dependence of the order parameter on the phonon frequency ($\omega$) is presented in Figure \ref{f3}. Specifically, plots therein depict the real and imaginary part of order parameters as obtained for the selected five representative values of temperature. For convenience, the Re$[\Delta(\omega)]$ and Im$[\Delta(\omega)]$ functions are imposed on the scaled Eliashberg function in the range $\omega \in \left< 0,\Omega_{max} \right>$.

As can be seen from the Figure \ref{f3}, the obtained Re$[\Delta(\omega)]$ and Im$[\Delta(\omega)]$ functions are directly related to the shape of the Eliashberg function, despite the selected temperature. In what follows, the order parameter takes  on the highest values in the $\omega$ region where $\alpha^{2}F\left(\omega\right)$ function also presents noticeable maxima. Nevertheless, from the physical point of view, the most significant are the values of the order parameter at low $\omega$ values, because they give the major contribution to the superconducting energy gap. It is also important to note that the imaginary part of the order parameter takes on the zero values in the range of $\omega \in \left< 0,50 \right>$ meV, suggesting no damping effects exist for this range of frequencies. Based on the results presented in Fig. \ref{f3}, the predicted values of the $\Delta_{g}$ is 14.24 meV, which is only slightly higher than the estimation made in terms of the imaginary axis calculations ($\Delta_{g} = 14.12$ meV).

To allow comparison between our predictions and other estimates it is convenient to determine the characteristic and dimensionless ratios, that appear in the Bardeen-Cooper-Schrieffer (BCS) theory of superconductivity \cite{bardeen1, bardeen2}. The first of such parameters is the one defined on the basis of the order parameter ($R_{\Delta}$). By using the order parameter value at the real axis the corresponding ratio can be written as:
\begin{equation}
\label{eq5}
R_{\Delta}\equiv\frac{2\Delta\left(0\right)}{k_{B}T_{c}}.
\end{equation}
In our study, the Eq. (\ref{eq5}) gives $R_{\Delta}=3.84$, and notably exceeds the BCS theory predictions of $R_{\Delta}=3.53$. 

\begin{figure}
\includegraphics[width=\columnwidth]{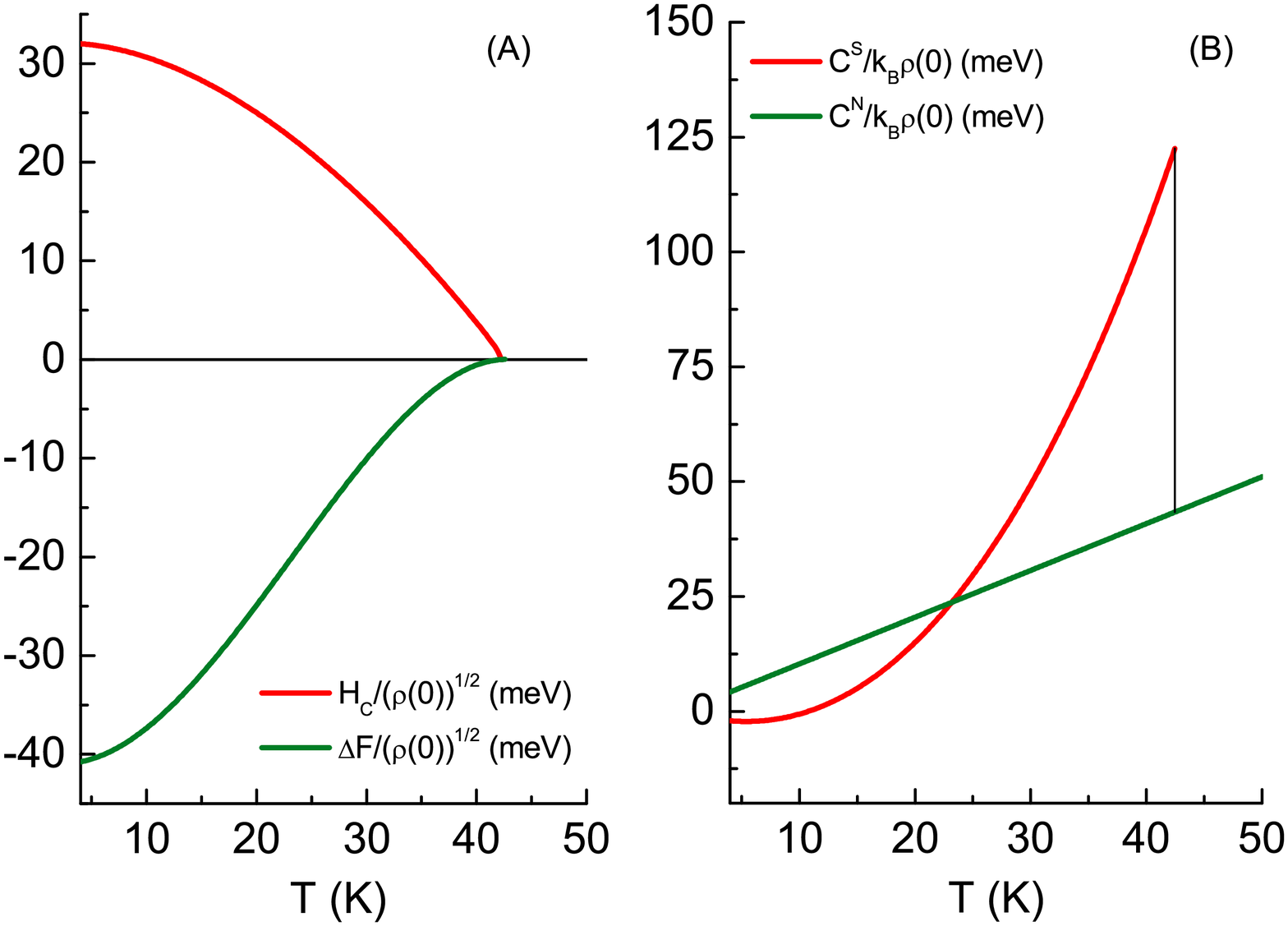}
\caption{(A - lower panel) The dependence of the free energy value on the temperature. (A - upper panel) The thermodynamic critical field as a function of the temperature. (B) The specific heat of the superconducting state and the normal state as a function of the temperature.} 
\label{f4}
\end{figure}

Similar dimensionless ratios can be calculated for the critical magnetic field ($H_{C}$), as well as the the specific heat difference between the superconducting and normal state ($\Delta C(T)$). However, first it is instructive to characterize above thermodynamic parameters. In what follows, the free energy difference between superconducting and normal state should be calculated in this respect. Herein, the $\Delta F$ parameter, normalized with respect to the electron density of states at the Fermi level ($\rho\left(0\right)$), is assumed in the following form \cite{bardeen3}:
\begin{eqnarray}
\label{eq6}
\frac{\Delta F}{\rho\left(0\right)}&=&-\frac{2\pi}{\beta}\sum_{n=1}^{M}
\left(\sqrt{\omega^{2}_{n}+\Delta^{2}_{n}}- \left|\omega_{n}\right|\right)\\ \nonumber
&\times&(Z^{S}_{n}-Z^{N}_{n}\frac{\left|\omega_{n}\right|}
{\sqrt{\omega^{2}_{n}+\Delta^{2}_{n}}}).
\end{eqnarray}  
In the above equation the $Z^{S}_{n}$ and $Z^{N}_{n}$ terms denote the wave function renormalization factors for the superconducting state and the normal state, respectively. The functional behavior of $\Delta F / \rho\left(0\right)$ parameter, in terms of the temperature, is depicted in the lower panel of Figure \ref{f4} (A). One can observe therein, that $\Delta F / \rho\left(0\right)$ function takes only negative values, suggesting thermodynamic stability of the discussed superconducting phase for $T\in \left<T_{0}, T_{c} \right>$. In addition, by using above results the critical magnetic field can be now computed from the formula:
\begin{equation}
\label{r11}
\frac{H_{c}}{\sqrt{\rho\left(0\right)}}=\sqrt{-8\pi\left[\Delta F/\rho\left(0\right)\right]}.
\end{equation}
We note, that the thermodynamic critical field strictly depends on the free energy difference and takes the positive values in the entire temperature range, as presented in the upper panel of Figure \ref{f4} (A). Moreover, in both cases, the $\Delta F$ and $H_{C}$ functions are equal to zero at the critical temperature. Hence, for $T>T_{c}$ the superconducting state disappears. To this end, the specific heat difference between superconducting state and the normal state ($\Delta C\equiv C^{S} - C^{N}$) is estimated on the basis of the formula:
\begin{equation}
\label{r12}
\frac{\Delta C\left(T\right)}{k_{B}\rho\left(0\right)}=-\frac{1}{\beta}\frac{d^{2}\left[\Delta F/\rho\left(0\right)\right]}{d\left(k_{B}T\right)^{2}},
\end{equation}
where the specific heat in the normal state is given by: $C^{N}(T)\slash{k_{\beta}\rho(0)}=\gamma\slash\beta$, assuming that the Sommerfeld constant is written as $\gamma=\frac{2}{3}\pi^{2}(1+\lambda)$. In Figure \ref{f4} (B) the temperature-dependent behavior of the specific heat of the superconducting state is presented in comparison to the normal specific heat. One can note a characteristic drop of the $C^{S}$ function at the critical temperature, which is physically relevant effect.

Based on the results given in Figure \ref{f4} (A) and (B) the two remaining thermodynamic ratios are determined according to the following relations:
\begin{equation}
\label{r13}
R_{c}\equiv\frac{\Delta C(T_{c})}{C^{N}(T_{c})},
\end{equation}
and
\begin{equation}
\label{r14}
R_{H}\equiv\frac{T_{c}C^{N}(T_{c})}{H^{2}_{C}(0)}.
\end{equation}
In case of HSe we obtain $R_{c}=1.83$ and $R_{H}=0.155$. Note, that in the framework of the BCS theory, the $R_{c}$ and $R_{H}$ parameters take the universal values equal to $R_{c}=1.43$ and $R_{H}=0.168$, respectively.

\section{Conclusions and summary}

In the present paper we have analyzed the thermodynamic properties of the HSe compund at $300$ GPa, to provide qualitative description of its superconducting phase. The numerical calculations were conducted within the Miglad-Eliashberg formalism for the specific value of the Coulomb pseudopotential equal to $\mu^*=0.1$.

During the analysis it was found that the critical temperature in HSe is relatively high ($T_{c}$=42.65 K) and surpass the magnesium diboride level by few Kelwins. High values of physical parameters were also found in terms of the superconducting energy gap and the effective mass of electrons. The employed theoretical formalism allowed us also to characterize such thermodynamic properties of the HSe as the free energy, the specific heat and the critical magnetic field. In reference to the obtained results we have estimated the characteristic thermodynamic ratios of the aforementioned parameters and found that their values strongly differ from the predictions of the BCS theory. As a result, we argue that the superconducting state is HSe is noticeably governed by the strong-coupling and retardation effects and cannot be properly described within the BCS theory. For convenience our numerical findings are summarized in Table \ref{tabela1}.

\begin{table}[h]
\centering
\caption{The most important thermodynamical parameters describing superconducting HSe under pressure of 300 GPa.}
\label{tabela1}
\begin{tabular}{| p{2cm} | p{2cm} | p{2cm}|}
\hline
\hline
Quantity & Unit & $HSe$ \\
\hline
$p$ & GPa & 300 \\
$max$ & - & 1100 \\
$\mu^{\star}$ & - & 0.1 \\
$\lambda$ & - & 0.79 \\
$\Omega_{max}$ & meV & 203.66 \\
\hline
$T_{c}$ & K & 42.65 \\
$T_{0}$ & K & 4 \\
$\Delta_{g}$ (Im axis) & meV & 14.12 \\
$\Delta_{g}$ (Re axis) & meV & 14.24 \\
\hline
$R_{\Delta}$ & - & 3.84 \\
$R_{c}$ & - & 1.83 \\
$R_{H}$ & - & 0.155 \\
\hline
\hline
\end{tabular}
\end{table}

\bibliography{bibliography}
\end{document}